\documentclass{aa}
\usepackage{txfonts}
\usepackage{subfig}

\begin{document}
	
\authorrunning{Qingshun Hu, Yu Zhang, Ali Esamdin et al.}
\titlerunning{The Layered Structure of Open Clusters}
	
\title{A detection of the layered structure of nearby open clusters}

\author{Qingshun Hu \inst{1,2} \and Yu Zhang \inst{1,2} \and Ali Esamdin \inst{1,2} \and Hong Wang \inst{1,2} \and Mingfeng Qin \inst{1,2}}

\institute{Xinjiang Astronomical Observatory, Chinese Academy of Sciences, No. 150, Science 1 Street, Urumqi, Xinjiang 830011, People's Republic of China, (\email{zhy@xao.ac.cn; aliyi@xao.ac.cn} \label{inst1}) \and University of Chinese Academy of Sciences, 19 Yuquan Road, Shijingshan District, Beijing 100049, People's Republic of China\label{inst2}
	}
	
\date{Received 7 June 2022; Accepted 30 January 2023}

\abstract{We applied the newly developed rose diagram overlay method to detect the layered structure of 88 nearby open clusters ($\leq$500~pc) on the three projections after the distance correction of their member stars, based on the catalog in literature. The results show that with the rose diagram overlay method, a total of 74 clusters in our sample have a layered structure, while the remaining clusters are without a clear layered structure. We for the first time defined the layered structure parameters for the sample clusters. Meanwhile, we found that the layered circle core area ($s$) has a strong positive correlation with the number of cluster members, while the kernel instability index ($\eta$) has a strong negative correlation with the number of cluster members. Our study provides a novel perspective for the detection of the layered structure of open clusters.}

	
\maketitle{}


\section{Introduction}

A stellar system, usually an open cluster, mostly forming in giant molecular clouds, is mainly located on the Galactic plane \citep{lada03}. There are two theories for the formation of open clusters, one of which is the theory of monolithic cluster formation \citep{lada84}. In this case, the cluster is first born in the central, densest region of the molecular cloud. After a violent expulsion of gas, the member stars in the dense cluster expand and eventually disperse into the field \citep{lada03}. The other theory is the mechanism of hierarchical star formation \citep{krui12}. In this framework, star formation in regions of high-density molecular clouds can produce gravitationally bound clusters. Stellar groups that formed in low-density regions have filamentary substructures that lose their bindings soon after the residual gas is expelled. Regardless of the cluster formation mechanism, it is inevitable that some imprints are left on the morphology of open clusters. Therefore, the study of their morphology can provide observational evidence of their formation mechanisms.

There have been many works on the morphology of open clusters for several decades. \citep[e.g.,][]{berg01, nila02, chen04, cart04, sant05, khar09, zhai17, dib18, hete19, tarr21}. In some studies, the typical shapes presented in open clusters are considered as core-corona structure \citep[e.g.,][]{chen04,zhai17,tarr21}. Numerical simulations performed by \citet{mcmi07}, \citet{alli09}, and \citet{moec09} indicated that star clusters form from the merging of substellar groups at a dynamically subviral stage, which to some extent supports the hierarchical star formation. Since the morphology of the merged subgroups may have some correlation with the morphology of clusters, a deeper study of the core-corona structure of clusters may verify the cluster formation mechanism from observations.

At present, the core-corona structure of open clusters has been studied in two-dimensional (2D) and three-dimensional (3D) spaces in the form of single sources \citep{zhan20,pang21}. However, there is a lack of statistical studies on the core-corona structure of open clusters, especially to investigate the intrinsic morphological properties and laws of the core-corona structure of open clusters. In our latest series of studies \citep{hu21a,hu21b}, we investigate the morphological properties and morphological stability of the core-corona structure of open clusters for the first time. 

In this work, we consider this core-corona structure as a layered structure for the first time. This layered structure of open clusters has been studied statistically in the framework of a 2D spherical coordinate system in previous works \citep{chen04,zhai17,hu21a,hu21b,tarr21}. However, the projection effect in 2D space may cause false positives for the layered structure of open clusters in 3D space. The questions are whether the layered structure of all open clusters is prevalent in 3D space, and which structural properties they have, statistically. These questions need to be addressed urgently. In the present work, we intend to detect the layered structure of open clusters in 3D space with a new method (rose diagram overlay) that we develop in this work and try to statistically analyze the properties of the layered structure in 3D space.

This paper is organized as follows. We describe in Sect.~2 the data and method we used in our study. In Sect.~3 we present some results. In Sect.~4 we discuss the robustness assessment of the correlations. Finally, Sect.~5 summarizes our results.

\section{Data and methods}

The sample in this paper was selected from the catalog of open clusters released by \citet{cant20a}, who provided the combined spatiokinematic-photometric membership of 1481 open clusters based on Gaia Data Release 2 (Gaia~DR2). The selected sample contains 88 open clusters that all pass the cuts of member' number (N)~$\geq$~50 and cluster center parallax ($\varpi$)~$\geq$~2. The constraint of N~$\geq$~50 can facilitate the depiction of the cluster morphology to some extent, while $\varpi$~$\geq$~2 allows the error in the distance correction of the sample cluster members to be reduced to about 6.3~pc, as pointed out by \citet{pang21}. The advantage of sample clusters is that they span a wide range of ages, from a few million years (young system) to several gigayears (old system), which meets the requirements of our study. In addition, we note that the members of the sample clusters adopted in this work are within G~$\leq$~18 \citep{cant20a}.

\subsection{Rose diagram overlay on 3D projection planes}

\label{Sec:rose}

A rose diagram usually consists of a certain number of sectors with the same angle, which can vary in radial extension radius. In short, it is a polar distribution diagram used to describe wind distribution and frequency and is named for its resemblance to a rose. It is also used to depict the distribution of velocity vectors \citep[e.g.,][their Fig~14]{mein21}. The rose diagram has the advantage of quantifying the characteristics of each sector. The 2D morphology of open clusters and their projected morphology in 3D space is irregular. At the same time, if these morphologies are divided into equal-angle sectors according to their centroids, the distribution of the member stars in these sectors is still irregular. Therefore, the morphology of open clusters is more suitable to be depicted by the rose diagram. We applied the rose diagram for the first time in this work to describe the morphology of open clusters.

Theoretically, it can be determined by visual inspection of its 3D spatial distribution of member stars whether the 3D spatial shape of a sample cluster has a layered structure. However, this approach cannot well describe or quantitatively study the layered structure of the sample cluster in 3D space.  Therefore, with the guidance of the rose diagram overlay method, we attempted to investigate whether the layered structure of sample clusters exists in 3D space.

\begin{figure*}
	\centering	
	\subfloat[]{
		\centering
		\includegraphics[width=130mm]{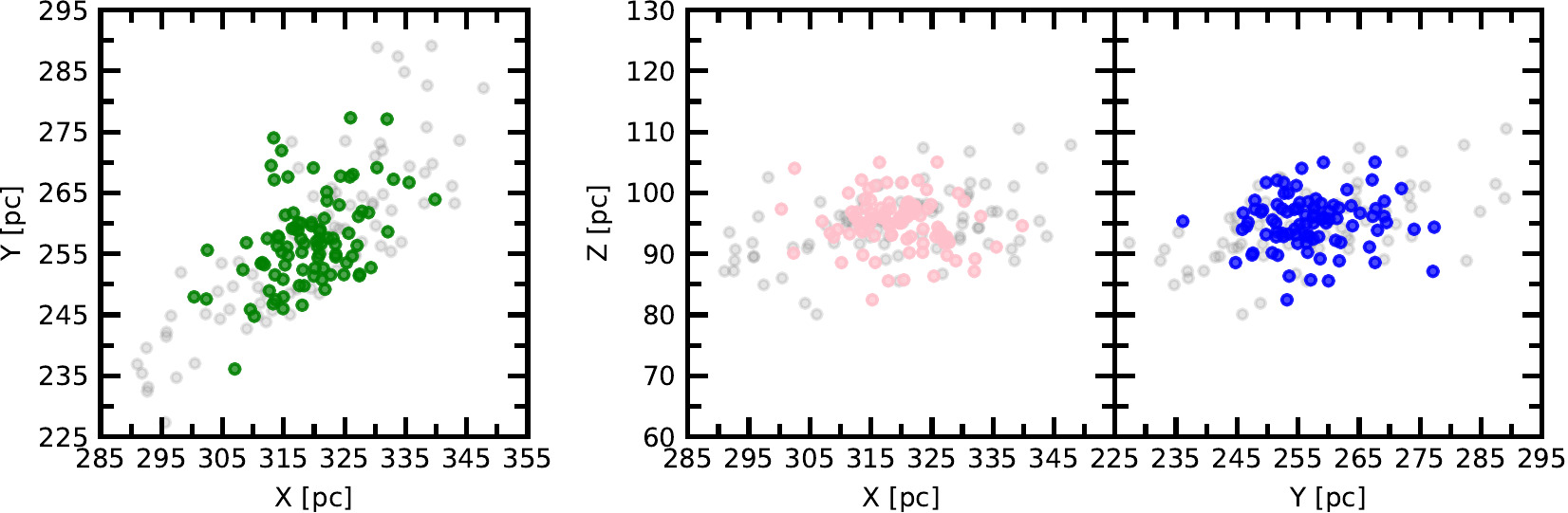}
	}\\
	\subfloat[]{
		\centering
		\includegraphics[width=130mm]{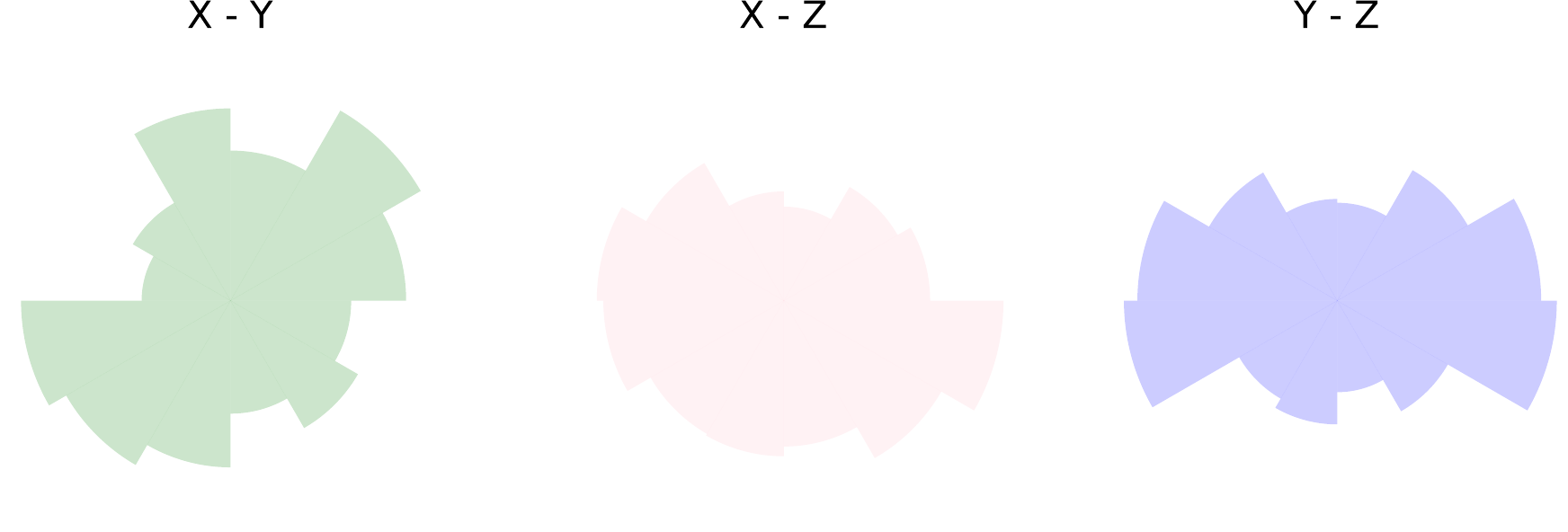}
	}
	
	\caption{(a) Members' distributions of the Gulliver~20 cluster on three projection planes after and before correcting distances. (b) Rose diagrams of the Gulliver~20 cluster on three projection planes after correcting distances. Distributions of stellar members of Gulliver~20 in the 3D projection planes as well as their rose diagrams in different projection planes. Upper panel: All colored dots in each panel represent the stellar members of the cluster (Gulliver~20) after distance correction. The gray dots in the background in each panel show the spatial distribution of all members of the cluster without distance correction. Bottom: Three different rose diagrams including the green rose diagram (X-Y plane), pink rose diagram (X-Z plane), and blue rose diagram (Y-Z plane).}
	\label{fig:Gulliver_20}
\end{figure*}

Generally, we present the morphologies of sample clusters on three different projections through the rose diagram and then overlaid these morphologies. Specifically, there are two steps to implementing the overlay ideal. Firstly, we can obtain the projection distributions of the members of each sample cluster on the X-Y, X-Z, and Y-Z planes; see Figure~\ref{fig:Gulliver_20}. Based on the distance correction described in Sect~\ref{Sec:Bayesian}, we calculated the heliocentric Cartesian coordinates\footnote{This refers to XYZ Cartesian coordinates centered on the Sun. The positive X-axis points from the projection of the position of the Sun onto the Galactic midplane toward the Galactic center. The positive Y-axis points in the direction of Galactic rotation, and the Z-axis is positive toward the Galactic north pole. The origin of the Cartesian heliocentric coordinate system is the Solar System barycenter.} (X, Y, Z) for each star member of the sample clusters by using the Python Astropy package \citep{astr13, astr18}. For the package, we adopted the default values for the Galactocentric coordinate frame, namely ICRS coordinates (RA, DEC) of the Galactic center = (266.4051$^{\circ}$, -28.936175$^{\circ}$), Galactocentric distance of the Sun = 8.122 kpc, and height of the Sun above the Galactic midplane = 20.8 pc. Figure~\ref{fig:Gulliver_20} shows the distribution (colored dots) of the Gulliver~20 members on the three projections. Meanwhile, the members (gray dots) of the cluster on the background show the distribution of the three projections before correcting their distances (for more details in Sect~\ref{Sec:Bayesian}).

Second, we started to construct the rose diagram of each sample cluster on different projection planes by using star counting and the median distance from the star to the center of the cluster. A 3$\sigma$ principle was applied to the distribution of members of the sample clusters on different planes before constructing a rose diagram. This typically was a range containing a mean plus or minus three standard deviations, which can remove some possible outliers in the rose diagram construction. We divided the distribution of the member stars of each sample cluster in each projection plane into 12 sectors, each with an angle of 30 degrees, as shown in the lower row (panel (b)) of Figure~\ref{fig:Gulliver_20}. The division into 12 sectors is a compromise between the number of member stars contained in a single sector and the extension of its radius. The radius (r$_{i}$) of each sector mainly depends on two parameters. The first parameter is the number (n$_{i}$) of member stars contained in each sector. The second is the median of the distance (d$_{i}$) from the member stars contained in each sector to the center of the sector. We recall that both covariates (n$_{i}$ and d$_{i}$) need to be normalized inside each sample cluster. They are normalized by the following equation:

\begin{equation}
\centering
n^{norm}_{i} = \frac{n_{i}}{n^{max}_{i}},  \qquad
d^{norm}_{i} = \frac{d_{i}}{d^{max}_{i}} 
\end{equation}

The relation between these two normalized covariates (n$^{norm}_{i}$ and d$^{norm}_{i}$) and the radius of each sector are shown in the following equation:

\begin{equation}
r_{i} = \sqrt{(n^{norm}_{i})^2 + (d^{norm}_{i})^2}
\label{equation2}
\end{equation}

r$_{i}$ includes r$^{XY}_{i}$, r$^{XZ}_{i}$, and r$^{YZ}_{i}$, with $i$ ranging from 1 to 12. According to the above expressions, the parameters (r$^{XY}_{i}$ or r$^{XZ}_{i}$ or r$^{YZ}_{i}$) of each projection plane were calculated. The rose diagrams of each sample cluster on different projection planes in 3D space can therefore be drawn. Finally, we overlaid the rose diagrams of each cluster on the three spatial projection planes to determine whether each sample cluster has a layered structure. 

In our work, we regarded the presence or absence of a circular inner core in the superimposed rose diagram of a sample cluster (r$_{min}$ $\neq$ 0 or r$_{min}$ = 0, r$_{min}$ = min[ min(r$^{XY}_{i}$),  min(r$^{XZ}_{i}$), min(r$^{YZ}_{i}$)]) as an indicator of whether the sample cluster had a layered structure. If there is a regular circle kernel (r$_{min}$ $\neq$ 0) in the superimposed rose diagram of the sample cluster, it can be determined that the sample cluster has a layered structure; if there is no regular circle kernel (r$_{min}$ = 0) in the superimposed rose diagram of the sample cluster, the sample cluster certainly has no layered structure detected by this method. 

\begin{figure}
	\centering
	\includegraphics[angle=0,width=50mm]{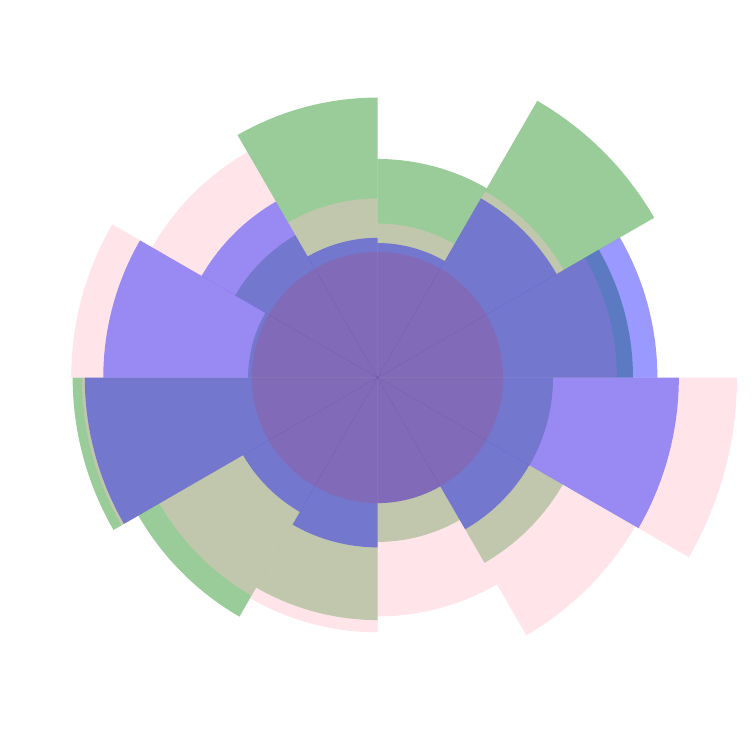}
	\caption{Overlaid rose diagram of the cluster (Gulliver~20). There are three different rose diagrams (see Sect.~\ref{Sec:rose} for more details about constructing them) on the projection planes in the heliocentric Cartesian coordinate system before overlaying, with the X-Y plane (green), X-Z plane (pink), and Y-Z plane (blue). The circle region in dark purple represents the circle core of the cluster in a layered structure, while the region in dark blue represents its irregular core.}
	\label{fig:Gulliver_20_position}
\end{figure}

\begin{figure*}
	\centering
	\includegraphics[angle=0,width=130mm]{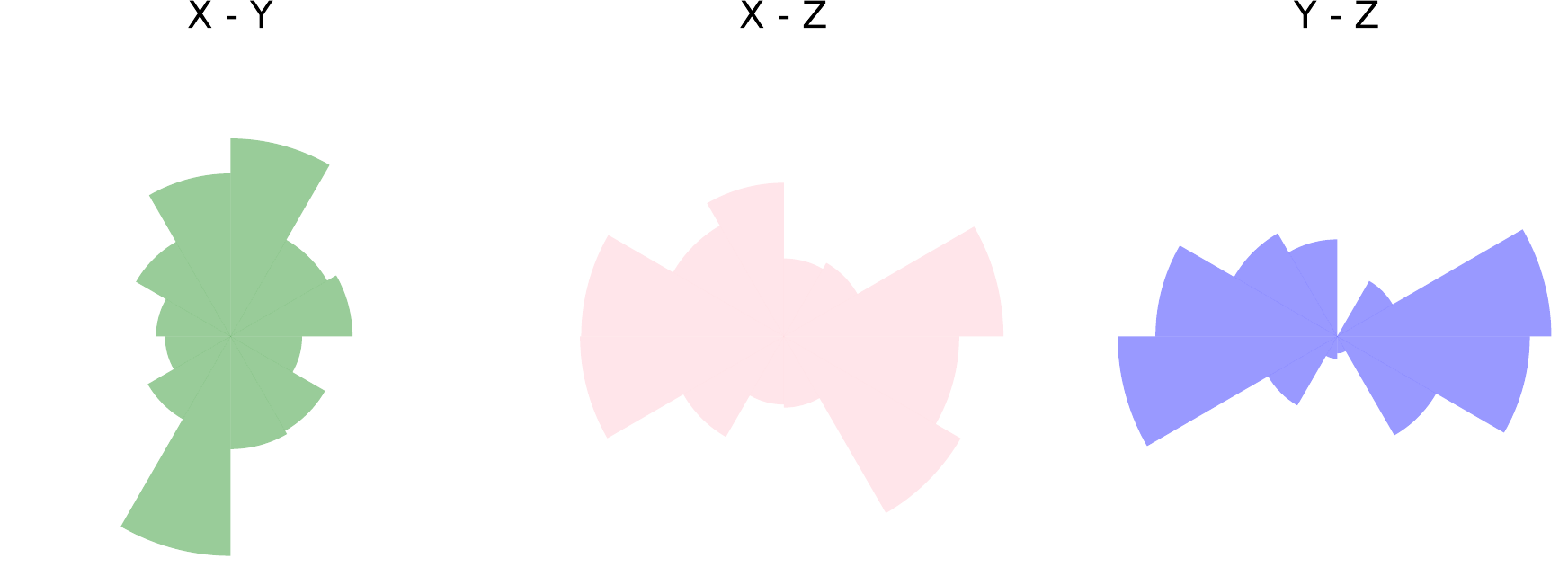}
	\caption{Rose diagrams of the cluster (BH~23) in different projection planes. The color information is consistent with panel (b) of Figure~\ref{fig:Gulliver_20}.}
	\label{fig:BH_23}
\end{figure*}

\begin{figure}
	\centering
	\includegraphics[angle=0,width=50mm]{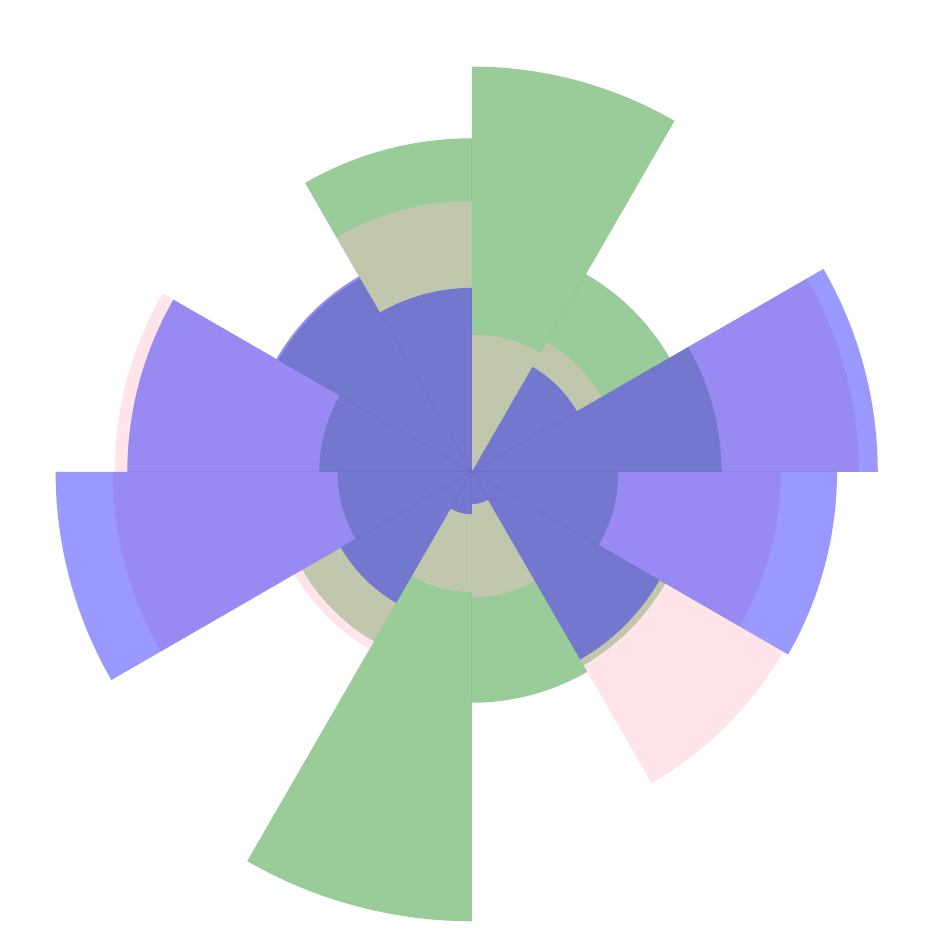}
	\caption{Overlaid rose diagram of the cluster (BH~23). The color information is the same as in Figure~\ref{fig:Gulliver_20_position}.}
	\label{fig:BH_23PPPolar}
\end{figure}

For example, we found that the Gulliver~20 cluster has a layered structure due to a regular circle kernel (the dark purple region) in its superimposed rose diagram; see Figure ~\ref{fig:Gulliver_20_position}. This is shown in panel (b) of Figure~\ref{fig:Gulliver_20} (r$_{min}$ $\neq$ 0). However, according to three different rose diagrams of the cluster BH~23, r$_{min}$ is equal to zero; see the Y-Z rose diagram of Figure ~\ref{fig:BH_23}. No regular circle kernel is found in its superimposed rose diagram, as shown in Figure ~\ref{fig:BH_23PPPolar}. Therefore, it does not have a layered structure.

\subsection{Correcting distances through Bayesian parallax inversion}

\label{Sec:Bayesian}

Several studies \citep[see, e.g.,][]{smit96, bail15, luri18, carr19, zhan20} have shown that the morphology of open clusters appears to be stretched along the line of sight. Because observational errors in measuring parallaxes are nonnegligible, their direct propagation to distances through inversion may yield an undesirable result. Similarly, there is such artificial elongation in directly inverting the DR~2 parallax, 1/$\varpi$. The errors in the parallax measurements $\Delta$$\varpi$ have a symmetric distribution function. After this reciprocation, the distance distribution becomes asymmetric. To alleviate this problem, we adopted the method introduced by \citet{bail15} and later adopted by \cite{pang21}. The distance correction procedure we closely followed is from \citet{carr19}; see their appendix. The distance correction problem is solved within a Bayesian framework. 


In this approach, the Bayesian theorem is adopted to estimate a posterior distribution that consists of a prior and likelihood function. Each star’s prior function includes two components, respectively, the members' density of clusters that shows a normal distribution and the distribution of field stars that is an exponentially decreasing distribution of density \citep{bail15}. These two components are combined by the membership probability as a weighting factor. We adopted a value of $\alpha$ to denote the probability of stellar membership provided by \citet{cant20a}, which denotes the cluster member term. And 1 - $\alpha$ is used to represent the field star. The standard deviation of the clustercentric distance of individual stars was set as the scale radius of the normal distribution. We calculated the likelihood function with the observed parallax and its error. The posterior mean is the corrected distance of each member. The detailed distance correction procedure can be referred to \cite{carr19} and \cite{pang21}. Moreover, we corrected the systematic offset of the measurement parallax by adding -0.029 mas \citep[][see their Figure 8]{lind18}. Figure~\ref{fig:Gulliver_20} shows the results obtained by applying this method. The distribution of the gray dots (denoting the Gulliver~20 member stars) in the figure indicates the elongated shape along the line of sight when there is no distance correction for its members. After distance correction, it presents a more compact distribution than the artificial elongation along the line of sight; see the colored dots in the three panels of this figure.

\section{Result} \label{result}

\subsection{Maximum stretch of sample clusters on three projection planes}

\begin{figure*}
	\centering
	\includegraphics[width=160mm]{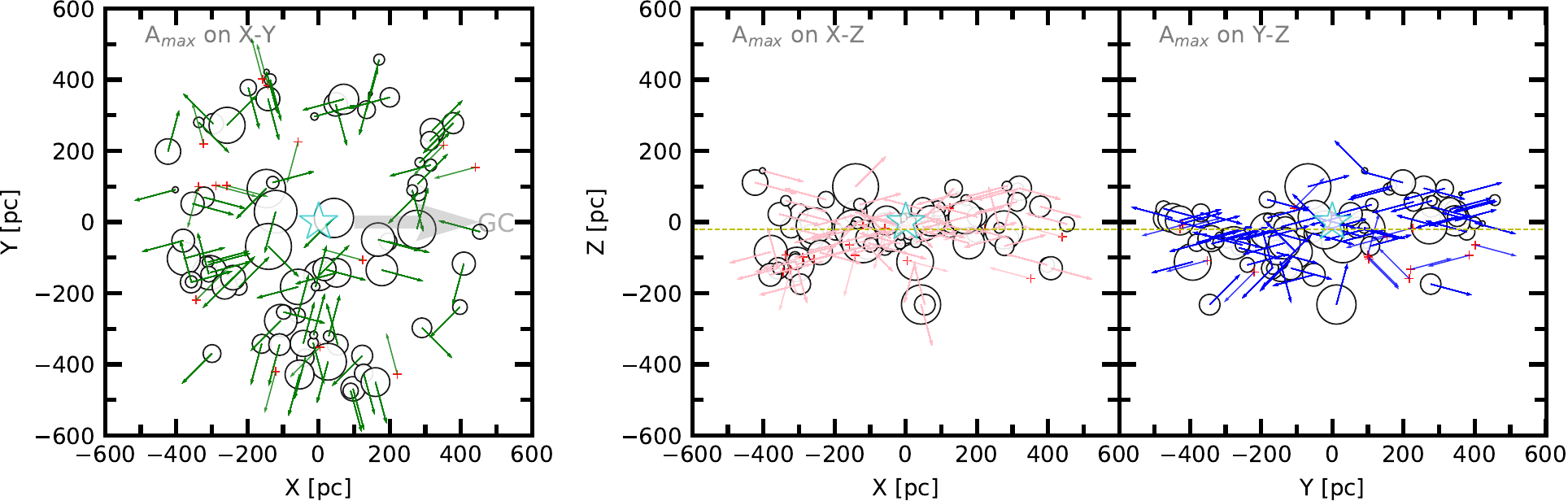}
	\caption{Simple morphological distribution of the sample clusters in the three projection planes. The size of open circles is proportional to the circle core area ($s$) of the sample clusters in each projection plane. All colored arrows mark the maximum stretch directions (A$_{max}$) of the sample clusters in each projection plane. The red crosses in each panel represent sample clusters without a layered structure. The open cyan pentagram in each panel denotes the Sun and the thick gray arrow in the left panel indicates the direction of the Galactic center. The dashed yellow line (Z = -20.8~pc) in the middle and right panel is designed as the Galactic midplane of the Milky Way. For more details, see Sect~\ref{Sec:rose}.}
	\label{fig:Sample_distribution}
	
\end{figure*}

Before superimposing the rose diagrams for each sample cluster, we obtained the rose diagrams of 88 sample clusters on different projection planes. We found that the rose diagrams of each sample cluster in three projection planes have a sector of maximum radius (max[r$_{i}$]). This sector of maximum radius seems to tell us the direction of maximum stretching of the sample cluster's morphology in each projection plane. The morphological stretching of the sample clusters on each projection plane can be attributed to specific external disturbances. To investigate the morphological maximum stretching pattern of the sample clusters, we statistically obtained the morphological maximum stretching directions of all the sample clusters on three different projections.

Figure~\ref{fig:Sample_distribution} shows the simplified distribution of the maximum stretch direction of the sample clusters on three different projection planes. In this figure, arrows of different colors are used to characterize the direction of maximum stretching of the morphology of the sample clusters on different projection planes. To distinguish sample clusters with or without a layered structure, we use open circles to represent sample clusters with a layered structure and crossess to denote sample clusters without a layered structure.

In the X-Y projection plane of Figure~\ref{fig:Sample_distribution}, some of the sample clusters in the direction of the Y-axis appear to be elongated along the direction of the Galactic differential rotation (usually perpendicular to the gray arrow pointing to the Galactic center in the left panel), which may indicate that the elongation of layered structures of these sample clusters (see the green arrows) in the current projection may be the result of the disturbance by external forces embedded in Galactic differential rotation. On the other hand, the direction of the arrows changes with increasing or decreasing values of the Y-axis. In addition, the elongation of most sample clusters on the X-Z and Y-Z projection planes is mostly parallel to the Galactic midplane, as shown in the median and right panels of Figure~\ref{fig:Sample_distribution}. This is broadly consistent with the findings of \citet{oort79} and \citet{berg01}.

\subsection{Layered structure parameters of sample clusters}

\label{parameters}

The layered structure of open clusters lays the foundation for understanding their morphological evolution. We obtained the overlaid rose diagrams of 88 sample clusters. In our sample, a total of 74 clusters with a layered structure were discovered, while the remaining clusters may lack a layered structure. In this subsection, we try to parameterize the layered structure of sample clusters and explore some possible potential patterns of their morphologies.

The area of the circular core of the layered structure comes to mind first. It represents the smallest regular nucleus of the cluster. For the same cluster, the larger its regular nucleus, the more significant or stable its layered structure. We defined it as the circle core area ($s$). The detailed equation involved is listed below,

\begin{equation}
s = \pi \cdot r_{min}^2 
\end{equation}

Except for the circular cores of sample clusters, there is an irregular nuclear region (e.g., the darkest colored area in Figure ~\ref{fig:Gulliver_20_position}) in the overlying rose diagram of each sample cluster. These irregular nuclear areas of the sample clusters can measure the instability of their nuclear regions. In this paper, we use $\Delta$$S$ to represent the irregular nuclear region of each sample cluster with the following equation:

\begin{equation}
\Delta S = \sum_{i=1}^{j} (\frac{\pi \cdot min(r_{i})^{2}}{12}) - s
\end{equation}

Based on the above two parameters, the instability index of the nuclear region of the sample clusters can be defined as the ratio of $\Delta$$S$ to $s$, as in Equation~\ref{e1},

\begin{equation}
\eta = \frac{\Delta S}{s} 
\label{e1}
\end{equation}

The s, $\Delta$ S, and $\eta$ are dimensionless, with $j$ = 12. The higher the instability index of the core regions of the sample clusters, the more unstable the layered structure of the clusters, and the weaker the cluster abilities to resist external perturbations.

\subsection{Correlations of the layered structure parameters of sample clusters with spatial position, number of members, and age}

\label{Sec:Correlations}

Theoretically, the layered structure of sample clusters should be attributed mainly to the external environment and self-evolution. The external environment may be different in different spatial positions. The position we wish to study is within the frame of heliocentric Cartesian coordinates. Therefore, we aim to explore the relation between the layered structure parameters of the sample clusters and their spatial positions.

According to the analysis, the instability index of the nuclear region ($\eta$) of the sample clusters depends weakly on their spatial position along the Y-axis, referring to its correlation parameters in Table~\ref{table:Correlation} (in which the correlations between other variables are listed as well), implying a possible nonhomogeneous external environment along the Y-axis in the vicinity of the solar neighborhood. We should note that this correlation appeared to disappear in a later series of correlation tests (for more details, see Sect~\ref{Sec:test}). Therefore, this conclusion needs to be treated with caution.

\begin{table*}[ht]\scriptsize
	\caption{Correlation information.}
	\centering
	
	\label{table:Correlation}
	\begin{tabular}{c c c c c c c}
		\hline\noalign{\smallskip}
		\hline\noalign{\smallskip}
		Variables & Pearson & $p$-value  & Pearson & $p$-value  & Pearson & $p$-value  \\
		& (\texttt{Med}) & (\texttt{Med}) &  (\texttt{Med+MAD}) & (\texttt{Med+MAD}) &  (\texttt{Med-MAD}) & (\texttt{Med-MAD}) \\
		(1) & (2) & (3) &   (4) & (5)&  (6) & (7) \\
		\hline\noalign{\smallskip}
		log(Number) vs. $s$  & 0.69 & 1.48~$\times$~$10^{-11}$   & 0.69 & 1.40~$\times$~$10^{-11}$ &  0.69 & 1.35~$\times$~$10^{-11}$ \\
		
		log(Number) vs. log($\eta$) & -0.64  & 7.22~$\times$~$10^{-10}$ & -0.65  & 4.03~$\times$~$10^{-10}$ &  -0.63  & 2.37~$\times$~$10^{-9}$ \\
		
		Y vs. log($\eta$)   & 0.27  & 1.85~$\times$~$10^{-2}$ &  0.24  & 3.89~$\times$~$10^{-2}$ & 0.23  & 4.58~$\times$~$10^{-2}$  \\
		\hline\noalign{\smallskip}
	\end{tabular}
	\tablefoot{Column~(1) represents the name of variables; Column~(2), (4), and (6) represent Pearson's $\rho$ correlation coefficients, and Column~(3), (5), and (7) represent probabilities for rejecting the null hypothesis that there is no correlation. The \texttt{Med} refers to the use of the median cluster-centric distance of the members. The \texttt{Med-MAD} and \texttt{Med+MAD} represent the use of the minimum and maximum distances within the median absolute deviation (MAD) range, respectively.}
	\flushleft
\end{table*}

In addition, we found no evidence that the layered structure parameters ($s$, $\Delta$$S$, and $\eta$) of the sample clusters are associated with their ages. However, we find that $s$ of the sample clusters has a strong positive correlation with the number of their members. A negative correlation between the $\eta$ and the number of sample cluster members was detected, however, implying that the fewer the member stars of the sample cluster, the worse the core instability of its layered structure. This also probably suggests that the system stability of the sample clusters depends on their ability to bind themselves.


\section{Discussion}

\label{Sec:test}

\subsection{Impact of misclassified member stars}

We acknowledge that cluster members that are classified are likely to be influenced by the members considered because stars that are misclassified as belonging to the cluster or missing members can change the median cluster-centric distance (also the median distance from members to the sector center), especially for clusters with a small number of members.

\begin{table*}[ht]\scriptsize
	\centering
	\caption{Correlation information under constraints of stellar magnitude}
	\label{table:magnitude}
	\begin{tabular}{ccccccccc}
		\hline\noalign{\smallskip}
		\hline\noalign{\smallskip}
		Variables & \multicolumn{2}{c}{G $\leq$ 18~mag} &  \multicolumn{2}{c}{G $\leq$ 17~mag} & \multicolumn{2}{c}{G $\leq$ 16~mag} &  \multicolumn{2}{c}{G $\leq$ 15~mag} \\
		\hline\noalign{\smallskip}
		& Pearson & $p$-value & Pearson & $p$-value & Pearson & $p$-value & Pearson & $p$-value \\
		(1) & (2) & (3) & (4) & (5) & (6) & (7) & (8) & (9) \\
		\hline\noalign{\smallskip}
		log(Number) vs. $s$  & 0.69 & 1.48~$\times$~$10^{-11}$  & 0.64 & 8.57~$\times$~$10^{-9}$ & 0.63 & 2.19~$\times$~$10^{-7}$  & 0.54 & 2.54~$\times$~$10^{-4}$ \\
		
		log(Number) vs. log($\eta$) & -0.64  & 7.22~$\times$~$10^{-10}$ & -0.65  & 4.59~$\times$~$10^{-9}$ & -0.62  & 3.32~$\times$~$10^{-7}$ & -0.45  & 2.45~$\times$~$10^{-3}$ \\
		
		Y vs. log($\eta$)  & 0.27  & 1.85~$\times$~$10^{-2}$ & 0.14  & 2.60~$\times$~$10^{-1}$ & 0.01  & 9.50~$\times$~$10^{-1}$ & 0.05  & 7.70~$\times$~$10^{-1}$ \\
		\hline\noalign{\smallskip}
	\end{tabular}
	\tablefoot{Column~(1) represents the name of variables; Column~(2), (4), (6), and (8) represent Pearson's $\rho$ correlation coefficients, and Column~(3), (5), (7) and (9) represent probabilities for rejecting the null hypothesis that there is no correlation.}
	\flushleft
\end{table*}

Therefore, we need to evaluate the impact of the objective fact mentioned above on the conclusions about all our correlations between different parameters in Sect~\ref{Sec:Correlations}. We tested the robustness of these correlations in this work through the median absolute deviation (MAD) of the median cluster-centric distance. The correlation coefficients and $p$-values of the correlations when considering the minimum (\texttt{Med-MAD}, representing the median cluster-centric distance of the members minus its median absolute deviation) and maximum (\texttt{Med+MAD}, referring to the median cluster-centric distance of the members plus its median absolute deviation) distances within the MAD range for the testing are complied in Table~\ref{table:Correlation}. Our correlations are clearly robust.

\subsection{Impact of completeness of the member stars}

\begin{table*}[ht]\scriptsize
	\centering
	\caption{Correlation information in the limit of member star probability}
	\label{table:Probability}
	\begin{tabular}{ccccccc}
		\hline\noalign{\smallskip}
		\hline\noalign{\smallskip}
		Variables &  \multicolumn{2}{c}{Probabilty $\geq$ 0.5} & \multicolumn{2}{c}{Probabilty $\geq$ 0.6} &  \multicolumn{2}{c}{Probabilty $\geq$ 0.7} \\
		\hline\noalign{\smallskip}
		& Pearson & $p$-value & Pearson & $p$-value & Pearson & $p$-value  \\
		(1) & (2) & (3) & (4) & (5) & (6) & (7)  \\
		\hline\noalign{\smallskip}
		log(Number) vs. $s$  & 0.67 & 6.23~$\times$~$10^{-11}$  & 0.67 & 2.65~$\times$~$10^{-10}$ & 0.67 & 3.67~$\times$~$10^{-10}$   \\
		
		log(Number) vs. log($\eta$) & -0.60  & 2.52~$\times$~$10^{-8}$ & -0.57  & 3.65~$\times$~$10^{-7}$ & -0.63  & 5.26~$\times$~$10^{-9}$ \\
		
		Y vs. log($\eta$)  & 0.24  & 4.40~$\times$~$10^{-2}$ & 0.23  & 5.80~$\times$~$10^{-2}$ & 0.27  & 2.30~$\times$~$10^{-2}$  \\
		\hline\noalign{\smallskip}
	\end{tabular}
	\tablefoot{Column~(1) represents the name of variables; Column~(2), (4), and (6) represent Pearson's $\rho$ correlation coefficients, and Column~(3), (5), and (7) represent probabilities for rejecting the null hypothesis that there is no correlation.}
	\flushleft
\end{table*}

In this paper, in addition to considering the effect of the misclassification of the sample cluster members on the correlations, we should also consider the effect of the completeness of the member list of the sample clusters on the correlations. Therefore, we considered the G-band magnitude of the member stars as a constraint. Because we lack a fainter list of member stars, G $\leq$ 18~mag, G $\leq$ 17~mag, G $\leq$ 16~mag, and G $\leq$ 15~mag were set as the conditions for the completeness test of the member list in this work.

We tested the correlations using the data of the member stars of the sample clusters under the G-band magnitude constraints. We found that by controlling the completeness of the member's list, some of the sample clusters with a small number of member stars changed their rose diagrams on the three projections, which also led to changes in their superimposed rose diagrams. However, for sample clusters with a high number of member stars, the layered structure was hardly affected.

With sample clusters with a layered structure, we evaluated the correlations under each constraint and found that the first two correlations remain very reliable; see Table~\ref{table:magnitude}. The table shows the results of all completeness tests. Then, the third correlation shows a loss of signal. The analysis reveals that the number of samples involved in assessing this correlation is too small.

\subsection{Effect of the purity of the member stars}

In our sample, the list of member stars contains a member probability parameter, which means that each member star of a cluster does not have an equal probability of being a true member of the cluster. For the members with a low membership probability, they are likely to be contaminants of cluster member stars. Therefore, we tested the effect of the purity of the cluster members on the correlations in this paper.
	
We set membership probabilities greater than and equal to 0.5, 0.6, and 0.7 as the limitations for the purity test of the member list. Since \citet{cant20} used cluster members with membership probabilities greater than and equal to 0.7 as the dependent data for assessing the basic parameters (e.g., distance, age, and interstellar reddening) of clusters, the conditions of this test are set to cut off at membership probabilities greater than and equal to 0.7.

We tested the correlations separately according to the probability constraints of member stars. The results of the tests are compiled in Table\ref{table:Probability}. The table shows that the purity of the membership star list of the sample clusters has almost no effect on the first two correlations, but has some effect on the third correlation. According to our analysis, the restriction of the membership probability causes a decrease in the number of member stars of the sample clusters. This effect causes a change in the rose diagram on the three projections for the sample clusters with a small number of member stars, and it also changes their superimposed rose diagrams, leading to a decrease in the number of sample clusters with a layered structure. This eventually makes the third correlation invalid. Although the number of the sample clusters with layered structure fluctuates under these limitations, the first two correlations remain significantly unaffected.

\section{Summary}

We detected and studied the layered structure of nearby open clusters for the first time in heliocentric Cartesian coordinates frame by overlaying a rose diagram after distance correction. Our conclusions are listed below. 

First, a total of 74 sample clusters with a layered structure were detected in our 88 sample clusters. Second, the member stars of the sample clusters have a strong positive correlation with their layered core area ($s$). Third, there is a strong negative correlation between the member stars of the sample clusters and the core instability index ($\eta$) of the cluster's layered structure.

As a pilot study of the layered structure of open clusters by decomposing projections in 3D space, we still need to acknowledge the small coverage of the spatial distributions of the current sample clusters. However, the layered structure of open clusters is thought to be reliable and has research value. This is because open clusters with a layered structure can indicate to some extent that they have a relatively stable structure and may have a relatively long evolutionary lifetime. Open clusters that do not have a layered structure, may be more susceptible to rapid disintegration by external perturbations. In order to study the layered structure of open clusters more accurately and systematically, we call on future researchers to select as many complete and reliable catalogs of member stars as possible instead of focusing only on the discovery of new clusters.

\begin{acknowledgements}
	We are grateful to an anonymous referee for valuable comments which have improved the paper significantly. This work is supported by the National Key R\&D program of China for Intergovernmental Scientific and Technological Innovation Cooperation Project under No. 2022YFE0126200, the Natural Science Foundation of Xinjiang Uygur Autonomous Region, No. 2022D01E86, the National Natural Science Foundation of China under grant U2031204,  and the science research grants from the China Manned Space Project with NO.CMS-CSST-2021-A08. We would also like to thank Ms. Chunli Feng for touching up the language of the article. This study has made use of the Gaia DR2, operated by the European Space Agency (ESA) space mission (Gaia). The Gaia archive website is \url{https://archives.esac.esa.int/gaia/}.

Software: Astropy \citep{astr13,astr18}, Scipy \citep{mill11}, TOPCAT \citep{tayl05}.

\end{acknowledgements}



\begin{appendix}

\section{Correlation of simulated open clusters}

Using the rose diagram superposition method, we detected most open clusters with a 3D spatial layered structure within 500 pc of the Sun. The layered structure parameters of these clusters are strongly correlated with the fundamental cluster parameters. In order to further verify the correlations in the main text, we would like to simulate open clusters with 3D spatial layered structures by primarily simulating the projection distribution of their member stars in 3D space.

\subsection{Mock open clusters}

We mainly simulated the spatial distribution of member stars of open clusters within 500 pc of the Sun based on a Monte Carlo method. We statistically extracted some parameters from 88 real sample clusters as a priori parameters for our modeling of simple simulated clusters. The relevant parameters are the number (N) of member stars of the real sample clusters, the fitted ellipsoidal ellipticity (E) of the member stars of the real sample clusters in 3D space, the aggregation ratio (A$_{R}$) of the member stars of the real sample clusters on the projection plane in 3D space, the proportion (P$_{G}$) of the G-band magnitude distribution of the member stars of the real sample clusters, and the proportion (P$_{P}$) of the probability distribution of the member stars of the real sample clusters.

Our detailed procedure for simulating open clusters is as follows: First, we randomly generated the number (n) of member stars for each simulated open cluster according to the mean (logN = 2.19) and variance (3$\sigma$ = 1.29) of the logarithm of N. Second, we randomly generated the member distribution of the simulated open cluster with a certain number (n) of member stars in the ellipsoidal range with ellipticity E, according to the statistical relations for the real sample clusters. The relations are as follows:  \begin{equation} 
		E = -0.126\left( \pm0.056\right) \times log10\left( n\right) + 1.059 \left(\pm 0.137\right) 
	\end{equation}
    \begin{equation} 
    	A_{R} = -0.060\left( \pm0.015\right) \times log10\left( n\right) + 0.271 \left(\pm 0.036\right) 
    \end{equation}

    $$ \left\{
    \begin{array}{rcl}
    	{n \leq 100 }      &          { E  \geq   0.806}   & {A_{R} = 0.150}\\
    	{100 < n \leq 500 }          & {0.806 > E \geq 0.718}  & {A_{R} = 0.130}\\
    	{500 < n \leq 1500 }     &      {0.718> E \geq 0.657}  & {A_{R} = 0.095}\\
    	{n > 1500 }      &        {E < 0.657} & {A_{R} = 0.080}
    \end{array} \right. $$ And we then randomly rotated the member distribution of the simulated cluster centered on its center to obtain a simulated open cluster with a random orientation. Third, we assigned G-band magnitudes to the members of each simulated open cluster according to the P$_{G}$ of the real sample clusters, $$ \left\{ \begin{array}{rcl}
		{G \leq 10 }     & {P_{G} = 0.06}\\
		{10 < G \leq 15 }     & {P_{G} = 0.37}\\
		{15 < G \leq 18 }     & {P_{G} = 0.57.}
	\end{array} \right. $$ Four, we assigned probabilities to the member stars of each simulated open cluster according to the P$_{P}$ of the real sample clusters.

\subsection{Verifying the correlations}

\begin{figure*}
	\centering	
	\subfloat[]{
		\centering
		\includegraphics[width=140mm]{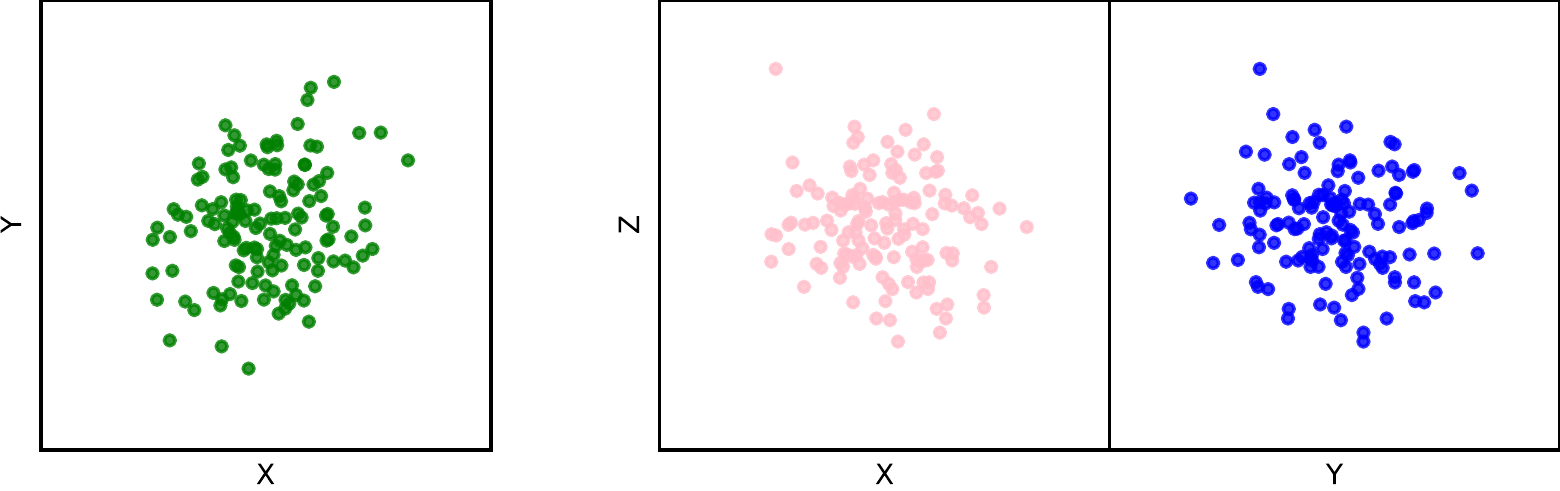}
	}\\
	\subfloat[]{
		\centering
		\includegraphics[width=140mm]{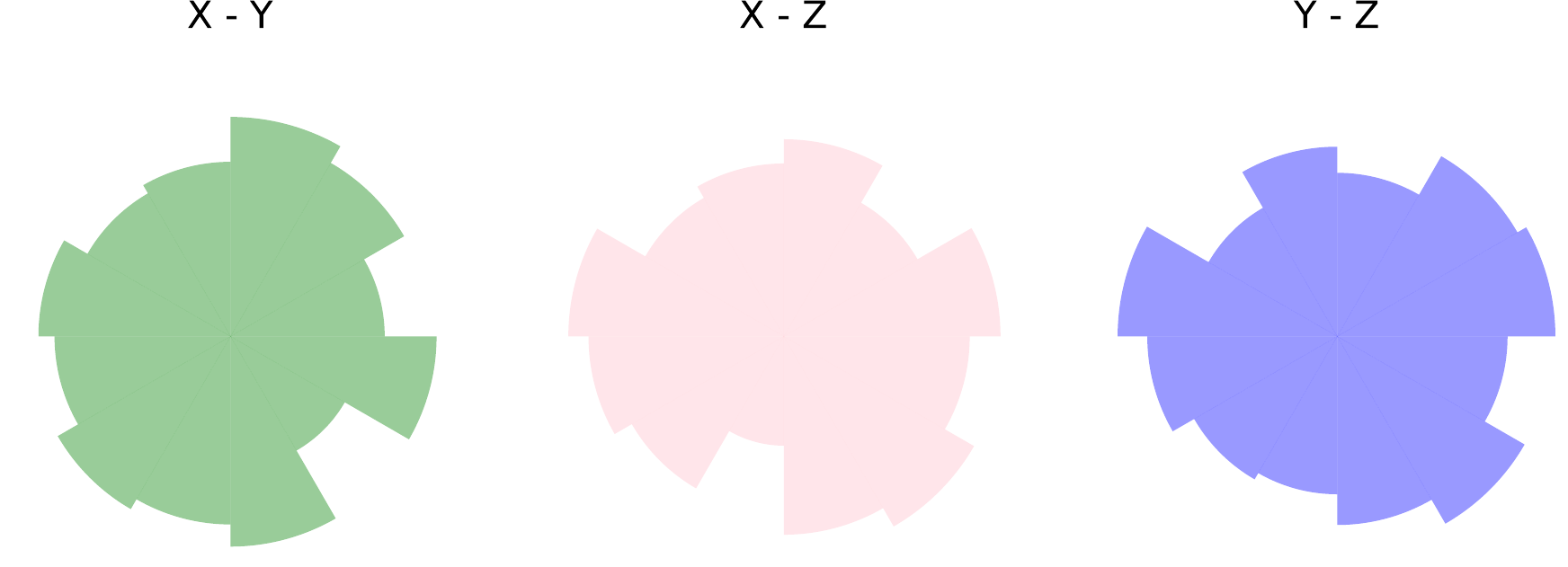}
	}
	
	\caption{(a) Members' distributions of the simulated cluster (with a layered structure) on three projection planes. (b) Rose diagrams of the simulated cluster (with a layered structure) on three projection planes. Distributions of stellar members of the simulated cluster (with a layered structure) in the 3D projection planes as well as their rose diagrams in different projection planes. The color information is the same as in panel (b) of Figure~\ref{fig:Gulliver_20}.}
	\label{fig:Mock_82}
\end{figure*}

\begin{figure*}
	\centering
	\includegraphics[angle=0,width=50mm]{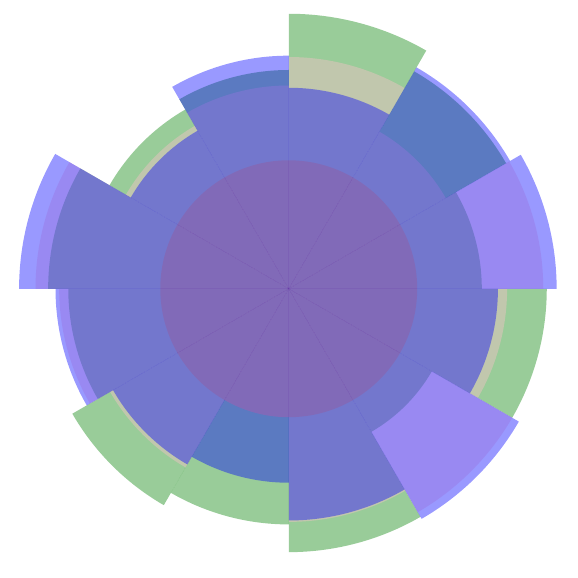}
	\caption{Overlaid rose diagram of the simulated cluster (with a layered structure). The color information is the same as in Figure~\ref{fig:Gulliver_20_position}.}
	\label{fig:Mock_layered_82}
\end{figure*}

\begin{figure*}
	\centering	
	\subfloat[]{
		\centering
		\includegraphics[width=140mm]{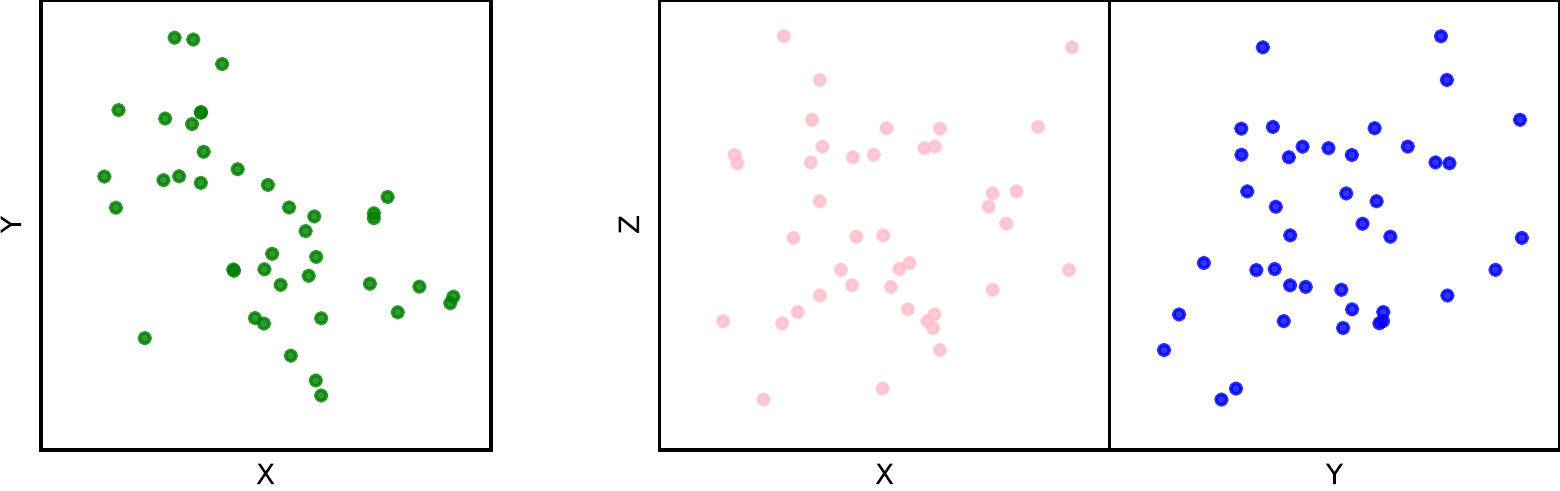}
	}\\
	\subfloat[]{
		\centering
		\includegraphics[width=140mm]{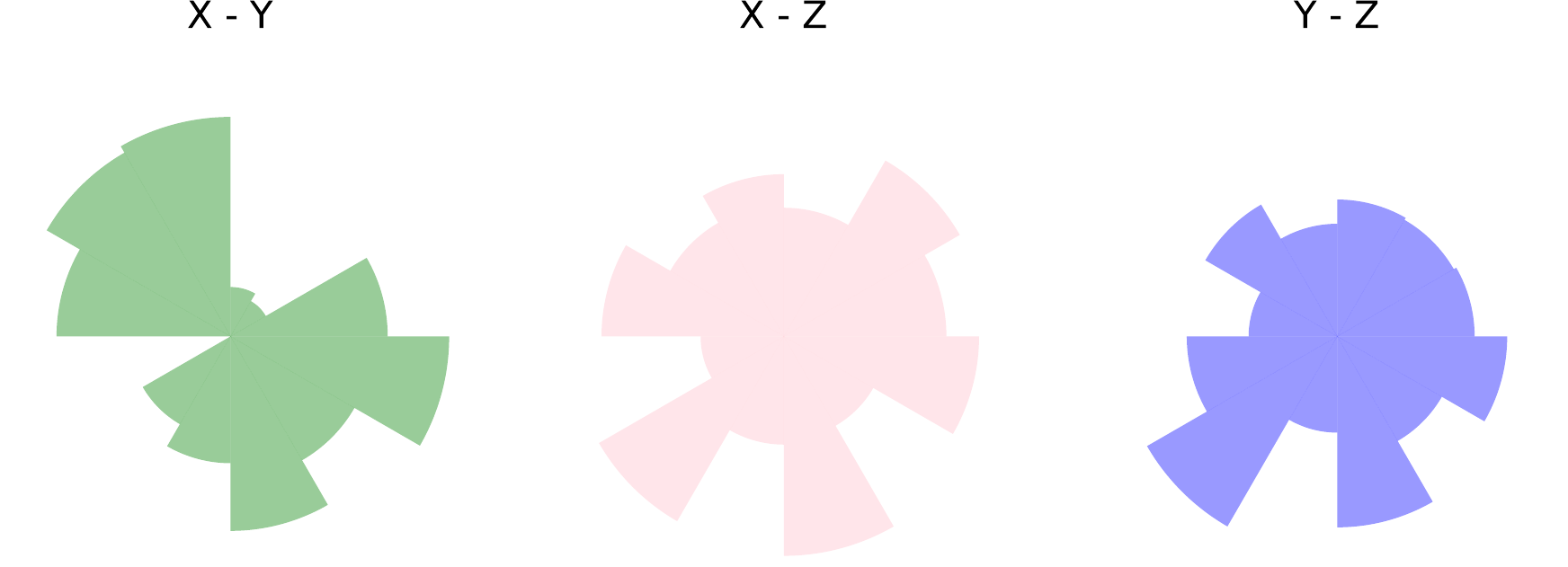}
	}
	
	\caption{(a) Members' distributions of the simulated cluster (without a layered structure) on three projection planes. (b) Rose diagrams of the simulated cluster (without a layered structure) on three projection planes. Distributions of stellar members of the simulated cluster (without a layered structure) in the 3D projection planes as well as their rose diagrams in different projection planes. The color information is the same as in panel (b) of Figure~\ref{fig:Gulliver_20}.}
	\label{fig:Mock_44}
\end{figure*}

\begin{figure*}
	\centering
	\includegraphics[angle=0,width=50mm]{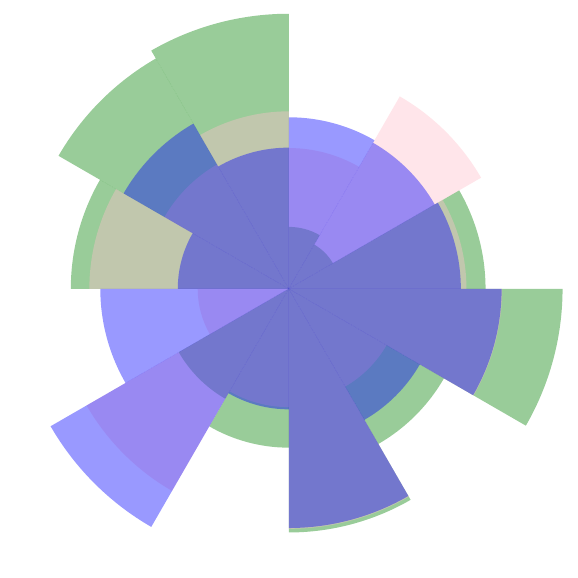}
	\caption{Overlaid rose diagram of the simulated cluster (without a layered structure). The color information is the same as in Figure~\ref{fig:Gulliver_20_position}.}
	\label{fig:Mock_layered_44}
\end{figure*}

With the above steps, our simulations yielded 440 simulated clusters, which is five times as many as the real sample clusters. This was done to improve the signal-to-noise ratio of potential correlations. Applying the rose diagram superposition method described in the main text to all of the simulated clusters, such as Figure ~\ref{fig:Mock_82}, Figure ~\ref{fig:Mock_layered_82}, Figure ~\ref{fig:Mock_44}, and Figure ~\ref{fig:Mock_layered_44}, we finally obtained the layered structure parameters of the simulated clusters statistically.

To investigate the relation between the basic cluster parameters for the simulated clusters and their layered structure parameters, we continued to use the analytical framework from the main text. We compiled the correlations between the number of member stars and the layered structure parameters (stability kernel ($s$) and instability index ($\eta$)) in Table~\ref{table:Correlation_mock}, without considering the influence of member stars by other factors and considering the effect of misclassification of member stars. This table shows that the correlations of the two sets of covariates remain consistent with those in the main text, and even their correlations are more robust, both when the effects of other factors are not taken into account and when member star misclassification is considered. This may be because the statistical sample of simulated clusters is larger than the actual cluster sample.

\begin{table*}[ht]\scriptsize
	\caption{Correlation information for simulated open clusters.}
	\centering
	
	\label{table:Correlation_mock}
	\begin{tabular}{c c c c c c c}
		\hline\noalign{\smallskip}
		\hline\noalign{\smallskip}
		Variables & Pearson & $p$-value  & Pearson & $p$-value  & Pearson & $p$-value  \\
		& (\texttt{Med}) & (\texttt{Med}) &  (\texttt{Med+MAD}) & (\texttt{Med+MAD}) &  (\texttt{Med-MAD}) & (\texttt{Med-MAD}) \\
		(1) & (2) & (3) &   (4) & (5)&  (6) & (7) \\
		\hline\noalign{\smallskip}
		log(Number) vs. $s$  & 0.91 & 5.84~$\times$~$10^{-141}$   & 0.92 & 7.54~$\times$~$10^{-148}$ &  0.91 & 6.28~$\times$~$10^{-134}$ \\
		
		log(Number) vs. log($\eta$) & -0.75  & 1.93~$\times$~$10^{-64}$ & -0.78  & 5.52~$\times$~$10^{-74}$ &  -0.73  & 7.12~$\times$~$10^{-61}$ \\
		\hline\noalign{\smallskip}
	\end{tabular}
	\tablefoot{Column~(1) represents the name of variables; Column~(2), (4), and (6) represent Pearson's $\rho$ correlation coefficients, and Column~(3), (5), and (7) represent probabilities for rejecting the null hypothesis that there is no correlation. The \texttt{Med} refers to the use of the median cluster-centric distance of the members. The \texttt{Med-MAD} and \texttt{Med+MAD} represent the use of the minimum and maximum distances within the median absolute deviation (MAD) range, respectively.}
	\flushleft
\end{table*}

In addition, we considered whether the completeness of member stars of the simulated cluster has an impact on the above two sets of correlations. Similarly, membership star magnitude restrictions can be equated to membership star completeness for correlation tests. According to the division in the main text, we continued to follow this criterion and counted the two sets of correlation parameters at G $\leq$ 18~mag, G $\leq$ 17~mag, G $\leq$ 16~mag, and G $\leq$ 15~mag, and the detailed parameters are shown in Table~\ref{table:magnitude_mock}. The number of simulated clusters with a layered structure decreases as the magnitude range shrinks when the two correlations are tested under the above completeness conditions, which leads to a decrease in the sample size of the simulated clusters involved in the correlation statistics. Finally, the reliability of the two correlations appears to be slightly weakened. However, the correlations of the two sets of covariates of the simulated clusters with a layered structure remain robust; see Table~\ref{table:magnitude_mock}. This may indicate that the intrinsic structural properties of clusters with layered structures mainly depend on the number of their member stars.

\begin{table*}[ht]\scriptsize
	\centering
	\caption{Correlation information under the constraints of stellar magnitude for the simulated open clusters}
	\label{table:magnitude_mock}
	\begin{tabular}{ccccccccc}
		\hline\noalign{\smallskip}
		\hline\noalign{\smallskip}
		Variables & \multicolumn{2}{c}{G $\leq$ 18~mag} &  \multicolumn{2}{c}{G $\leq$ 17~mag} & \multicolumn{2}{c}{G $\leq$ 16~mag} &  \multicolumn{2}{c}{G $\leq$ 15~mag} \\
		\hline\noalign{\smallskip}
		& Pearson & $p$-value & Pearson & $p$-value & Pearson & $p$-value & Pearson & $p$-value \\
		(1) & (2) & (3) & (4) & (5) & (6) & (7) & (8) & (9) \\
		\hline\noalign{\smallskip}
		log(Number) vs. $s$  & 0.91 & 5.84~$\times$~$10^{-141}$  & 0.91 & 1.50~$\times$~$10^{-134}$ & 0.89 & 3.03~$\times$~$10^{-109}$  & 0.87 & 1.52~$\times$~$10^{-90}$ \\
		
		log(Number) vs. log($\eta$) & -0.75  & 1.93~$\times$~$10^{-64}$ & -0.75  & 2.32~$\times$~$10^{-61}$ & -0.69  & 1.17~$\times$~$10^{-47}$ & -0.70  & 5.68~$\times$~$10^{-44}$ \\
		\hline\noalign{\smallskip}
	\end{tabular}
	\tablefoot{Column~(1) represents the name of variables; Column~(2), (4), (6), and (8) represent Pearson's $\rho$ correlation coefficients, and Column~(3), (5), (7) and (9) represent probabilities for rejecting the null hypothesis that there is no correlation.}
	\flushleft
\end{table*}

Finally, we considered the effect of the member star probabilities on the two sets of correlations of the simulated clusters. Referring to the main text, we set three conditional restrictions, probability $\geq$ 0.5, probability $\geq$ 0.6, and probability $\geq$ 0.7, respectively, and then carried out correlation tests of the two sets of covariates of the simulated clusters under these three conditions. The results of the correlation tests are included in Table~\ref{table:Probability_mock}. We find that there is no regular effect of the member star probability on the correlation of the two covariates, which produces a small fluctuation in the values of the two correlations, but does not cause a change in or disappearance of the statistically linear correlation of the simulated clusters with a layered structure.

\begin{table*}[ht]\scriptsize
	\centering
	\caption{Correlation information in the limit of member star probability for the simulated open clusters}
	\label{table:Probability_mock}
	\begin{tabular}{ccccccc}
		\hline\noalign{\smallskip}
		\hline\noalign{\smallskip}
		Variables &  \multicolumn{2}{c}{Probabilty $\geq$ 0.5} & \multicolumn{2}{c}{Probabilty $\geq$ 0.6} &  \multicolumn{2}{c}{Probabilty $\geq$ 0.7} \\
		\hline\noalign{\smallskip}
		& Pearson & $p$-value & Pearson & $p$-value & Pearson & $p$-value  \\
		(1) & (2) & (3) & (4) & (5) & (6) & (7)  \\
		\hline\noalign{\smallskip}
		log(Number) vs. $s$  & 0.91 & 5.68~$\times$~$10^{-137}$  & 0.91 & 2.55~$\times$~$10^{-135}$ & 0.91 & 2.27~$\times$~$10^{-130}$   \\
		
		log(Number) vs. log($\eta$) & -0.75  & 1.46~$\times$~$10^{-65}$ & -0.77  & 8.05~$\times$~$10^{-70}$ & -0.75  & 1.63~$\times$~$10^{-64}$ \\
		\hline\noalign{\smallskip}
	\end{tabular}
	\tablefoot{Column~(1) represents the name of variables; Column~(2), (4), and (6) represent Pearson's $\rho$ correlation coefficients, and Column~(3), (5), and (7) represent probabilities for rejecting the null hypothesis that there is no correlation.}
	\flushleft
\end{table*}

It should be noted that this simulation of open clusters was conducted to meet the statistical requirements. We only generated the 3D spatial distribution of the member stars of 440 simulated open clusters randomly based on the Monte Carlo idea, using the statistical parameters of the actual sample open clusters as reference. Complex simulations were not considered, and the G-band magnitudes of the member stars were not included for magnitudes fainter than 18, the coordinates of the simulated clusters in the Galactic disk were not considered, and the Cartesian coordinates of the simulated cluster members in the Galactic frame are not given. This is because we only need to consider the coordinates of each member in 3D relative to the cluster center, which allows us to carry out the subsequent layered structure detection and correlation tests. Overall, this simulation is based on the statistical parameters of the actual sample clusters.

\begin{table*}[ht]\scriptsize
	\centering
	\caption{The layered structure parameters of the sample clusters}
	\begin{tabular}{lcccccc}
		\hline\noalign{\smallskip}
		Cluster & $s$ & $\Delta$ S  & $\eta$ &  A\_xy &  A\_xz  &  A\_yz \\
		\hline\noalign{\smallskip}
		- & - &  - &  - &  degree &  degree  &  degree \\
		\hline\noalign{\smallskip}
		NGC$\_$2547 & 0.379 & 0.448 & 1.182 & 255.0 & 195.0 & 15.0 \\
		NGC$\_$3532 & 1.037 & 0.627 & 0.604 & 285.0 & 15.0 & 165.0 \\
		NGC$\_$1333 & 0.000 & 0.055 & 9999 & 345.0 & 15.0 & 315.0 \\
		Blanco$\_$1 & 1.933 & 1.240 & 0.642 & 225.0 & 285.0 & 75.0 \\
		UPK$\_$305 & 0.000 & 0.791 & 9999 & 105.0 & 255.0 & 15.0 \\
		NGC$\_$6633 & 0.440 & 1.155 & 2.622 & 45.0 & 195.0 & 195.0 \\
		Roslund$\_$6 & 1.137 & 1.270 & 1.117 & 195.0 & 345.0 & 15.0 \\
		Collinder$\_$135 & 1.328 & 1.013 & 0.763 & 225.0 & 345.0 & 165.0 \\
		Stock$\_$12 & 0.000 & 1.329 & 9999 & 105.0 & 15.0 & 345.0 \\
		UPK$\_$545 & 0.147 & 0.980 & 6.664 & 75.0 & 15.0 & 15.0 \\
		UBC$\_$7 & 0.260 & 1.167 & 4.495 & 345.0 & 195.0 & 165.0 \\
		BH$\_$164 & 0.491 & 0.393 & 0.800 & 315.0 & 165.0 & 15.0 \\
		Teutsch$\_$35 & 0.149 & 1.074 & 7.221 & 255.0 & 15.0 & 195.0 \\
		UPK$\_$292 & 0.135 & 0.756 & 5.608 & 315.0 & 345.0 & 195.0 \\
		Melotte$\_$22 & 2.292 & 0.910 & 0.397 & 255.0 & 315.0 & 15.0 \\
		L$\_$1641S & 0.000 & 0.565 & 9999 & 45.0 & 15.0 & 225.0 \\
		Gulliver$\_$6 & 0.579 & 1.367 & 2.360 & 15.0 & 195.0 & 225.0 \\
		IC$\_$4756 & 0.561 & 1.106 & 1.971 & 225.0 & 345.0 & 345.0 \\
		RSG$\_$5 & 0.691 & 1.032 & 1.494 & 285.0 & 165.0 & 15.0 \\
		UPK$\_$606 & 0.000 & 1.059 & 9999 & 165.0 & 195.0 & 45.0 \\
		UPK$\_$20 & 0.000 & 1.243 & 9999 & 195.0 & 165.0 & 195.0 \\
		UBC$\_$8 & 0.814 & 0.896 & 1.102 & 75.0 & 345.0 & 165.0 \\
		Pozzo$\_$1 & 0.854 & 0.368 & 0.431 & 75.0 & 135.0 & 15.0 \\
		Collinder$\_$69 & 1.373 & 0.995 & 0.725 & 15.0 & 195.0 & 135.0 \\
		Platais$\_$3 & 0.194 & 0.977 & 5.034 & 15.0 & 15.0 & 195.0 \\
		UPK$\_$422 & 0.755 & 0.959 & 1.271 & 225.0 & 345.0 & 345.0 \\
		Mamajek$\_$4 & 0.667 & 0.867 & 1.300 & 255.0 & 345.0 & 345.0 \\
		UPK$\_$640 & 1.396 & 0.914 & 0.654 & 345.0 & 195.0 & 345.0 \\
		RSG$\_$7 & 0.179 & 0.664 & 3.718 & 285.0 & 225.0 & 345.0 \\
		NGC$\_$2422 & 0.401 & 0.294 & 0.734 & 225.0 & 345.0 & 345.0 \\
		UPK$\_$540 & 0.000 & 1.082 & 9999 & 225.0 & 345.0 & 195.0 \\
		Alessi$\_$10 & 0.000 & 0.929 & 9999 & 225.0 & 165.0 & 135.0 \\
		Alessi$\_$5 & 0.559 & 0.741 & 1.326 & 225.0 & 195.0 & 15.0 \\
		Platais$\_$8 & 0.673 & 0.859 & 1.276 & 345.0 & 165.0 & 195.0 \\
		Alessi$\_$24 & 0.272 & 1.013 & 3.724 & 225.0 & 165.0 & 15.0 \\
		NGC$\_$2451A & 1.568 & 0.955 & 0.609 & 195.0 & 195.0 & 345.0 \\
		UPK$\_$535 & 0.082 & 1.353 & 16.510 & 75.0 & 165.0 & 345.0 \\
		UPK$\_$350 & 0.048 & 0.870 & 18.164 & 195.0 & 345.0 & 135.0 \\
		NGC$\_$2451B & 0.592 & 0.902 & 1.525 & 255.0 & 15.0 & 15.0 \\
		UBC$\_$17b & 0.182 & 0.981 & 5.396 & 105.0 & 105.0 & 105.0 \\
		Collinder$\_$350 & 0.182 & 1.249 & 6.854 & 195.0 & 195.0 & 195.0 \\
		UBC$\_$32 & 0.147 & 0.436 & 2.972 & 285.0 & 105.0 & 345.0 \\
		Melotte$\_$20 & 1.852 & 0.885 & 0.478 & 225.0 & 195.0 & 195.0 \\
		ASCC$\_$19 & 1.481 & 0.689 & 0.465 & 15.0 & 15.0 & 45.0 \\
		IC$\_$2602 & 1.027 & 1.495 & 1.456 & 195.0 & 345.0 & 15.0 \\
		ASCC$\_$21 & 0.156 & 1.365 & 8.764 & 15.0 & 15.0 & 225.0 \\
		ASCC$\_$16 & 0.320 & 1.155 & 3.609 & 15.0 & 195.0 & 225.0 \\
		RSG$\_$1 & 0.540 & 1.393 & 2.581 & 345.0 & 345.0 & 345.0 \\
		NGC$\_$752 & 0.533 & 0.767 & 1.438 & 135.0 & 195.0 & 345.0 \\
		UBC$\_$31 & 0.000 & 1.153 & 9999 & 255.0 & 285.0 & 75.0 \\
		Ruprecht$\_$147 & 0.459 & 1.216 & 2.650 & 45.0 & 15.0 & 15.0 \\
		Stock$\_$2 & 1.629 & 1.094 & 0.671 & 45.0 & 15.0 & 15.0 \\
		NGC$\_$1901 & 0.547 & 1.241 & 2.268 & 105.0 & 105.0 & 45.0 \\
		NGC$\_$2632 & 2.645 & 0.771 & 0.291 & 315.0 & 45.0 & 15.0 \\
		ASCC$\_$101 & 0.019 & 1.184 & 61.567 & 255.0 & 195.0 & 15.0 \\
		Gulliver$\_$20 & 0.742 & 0.815 & 1.098 & 45.0 & 345.0 & 345.0 \\
		Ruprecht$\_$98 & 0.000 & 1.205 & 9999 & 105.0 & 195.0 & 15.0 \\
		Stock$\_$1 & 0.474 & 1.245 & 2.625 & 195.0 & 165.0 & 195.0 \\
		BH$\_$23 & 0.000 & 0.768 & 9999 & 255.0 & 15.0 & 195.0 \\
		IC$\_$348 & 0.000 & 0.268 & 9999 & 345.0 & 195.0 & 315.0 \\
		ASCC$\_$58 & 0.742 & 1.163 & 1.568 & 285.0 & 195.0 & 345.0 \\
		NGC$\_$1662 & 0.652 & 0.955 & 1.465 & 195.0 & 15.0 & 195.0 \\
		NGC$\_$2516 & 1.696 & 0.391 & 0.230 & 255.0 & 255.0 & 195.0 \\
		ASCC$\_$127 & 0.711 & 0.896 & 1.260 & 285.0 & 345.0 & 15.0 \\
		UPK$\_$612 & 1.243 & 0.395 & 0.318 & 345.0 & 165.0 & 195.0 \\
		IC$\_$4665 & 0.116 & 1.184 & 10.177 & 45.0 & 195.0 & 195.0 \\
		Platais$\_$9 & 0.098 & 0.411 & 4.196 & 75.0 & 315.0 & 15.0 \\
		IC$\_$2391 & 1.141 & 1.110 & 0.973 & 45.0 & 225.0 & 195.0 \\
		Aveni$\_$Hunter$\_$1 & 0.000 & 1.019 & 9999 & 105.0 & 255.0 & 195.0 \\
		NGC$\_$2232 & 0.785 & 1.264 & 1.610 & 45.0 & 15.0 & 15.0 \\
		ASCC$\_$123 & 0.000 & 1.248 & 9999 & 255.0 & 255.0 & 195.0 \\
		RSG$\_$8 & 0.038 & 0.297 & 7.894 & 285.0 & 15.0 & 165.0 \\
		Collinder$\_$140 & 0.454 & 0.580 & 1.277 & 255.0 & 195.0 & 195.0 \\
		Alessi$\_$3 & 0.280 & 1.086 & 3.873 & 75.0 & 345.0 & 345.0 \\
		Alessi$\_$9 & 0.381 & 1.080 & 2.837 & 345.0 & 15.0 & 345.0 \\
		ASCC$\_$41 & 0.336 & 0.877 & 2.610 & 45.0 & 165.0 & 345.0 \\
		Alessi$\_$20 & 0.339 & 0.629 & 1.856 & 285.0 & 195.0 & 165.0 \\
		NGC$\_$6405 & 0.281 & 0.884 & 3.151 & 165.0 & 165.0 & 135.0 \\
		UPK$\_$533 & 0.137 & 0.669 & 4.892 & 285.0 & 165.0 & 15.0 \\
		Stephenson$\_$1 & 0.397 & 1.211 & 3.048 & 75.0 & 225.0 & 195.0 \\
		Trumpler$\_$10 & 1.051 & 1.425 & 1.355 & 255.0 & 345.0 & 15.0 \\
		NGC$\_$7092 & 0.073 & 1.651 & 22.607 & 15.0 & 165.0 & 165.0 \\
		NGC$\_$6475 & 1.809 & 0.789 & 0.436 & 75.0 & 75.0 & 15.0 \\
		NGC$\_$3228 & 0.296 & 1.233 & 4.165 & 285.0 & 15.0 & 345.0 \\
		Stock$\_$10 & 0.671 & 1.110 & 1.654 & 345.0 & 345.0 & 195.0 \\
		ASCC$\_$99 & 0.288 & 1.650 & 5.721 & 225.0 & 195.0 & 165.0 \\
		BH$\_$99 & 0.430 & 0.810 & 1.885 & 285.0 & 165.0 & 15.0 \\
		UBC$\_$17a & 0.313 & 1.063 & 3.401 & 105.0 & 285.0 & 285.0 \\
		\hline\noalign{\smallskip}
	\end{tabular}
	\tablefoot{A\_xy, A\_xz, and A\_yz represent the maximum stretch directions of the sample clusters in the X-Y plane, X-Z plane, and Y-Z plane, respectively. $s$ = 0 or $\eta$ = 9999 indicates the sample cluster without a layered structure.}
	\flushleft
\end{table*}

\end{appendix}


\begin{thebibliography}{}
	
	
	\bibitem[Allison et al.(2009)]{alli09} Allison, R.~J., Goodwin, S.~P., Parker, R.~J., et al.\ 2009, \apjl, 700, L99. doi:10.1088/0004-637X/700/2/L99
	
	\bibitem[Astropy Collaboration et al.(2013)]{astr13} Astropy Collaboration, Robitaille, T.~P., Tollerud, E.~J., et al.\ 2013, \aap, 558, A33. doi:10.1051/0004-6361/201322068
	
	\bibitem[Astropy Collaboration et al.(2018)]{astr18} Astropy Collaboration, Price-Whelan, A.~M., Sip{\H{o}}cz, B.~M., et al.\ 2018, \aj, 156, 123. doi:10.3847/1538-3881/aabc4f
	
	\bibitem[Bailer-Jones(2015)]{bail15} Bailer-Jones, C.~A.~L.\ 2015, \pasp, 127, 994. doi:10.1086/683116
	
	\bibitem[Bergond et al.(2001)]{berg01} Bergond, G., Leon, S., \& Guibert, J.\ 2001, \aap, 377, 462.
	doi:10.1051/0004-6361:20011043
	
	\bibitem[Bovy(2015)]{bovy15} Bovy, J.\ 2015, \apjs, 216, 29. doi:10.1088/0067-0049/216/2/29
	
	
	\bibitem[Cartwright \& Whitworth(2004)]{cart04} Cartwright, A. \& Whitworth, A.~P.\ 2004, \mnras, 348, 589. doi:10.1111/j.1365-2966.2004.07360.x
	
	\bibitem[Carrera et al.(2019)]{carr19} Carrera, R., Pasquato, M., Vallenari, A., et al.\ 2019, \aap, 627, A119. doi:10.1051/0004-6361/201935599
	
	\bibitem[Cantat-Gaudin et al.(2020)]{cant20} Cantat-Gaudin, T., Anders, F., Castro-Ginard, A., et al.\ 2020, \aap, 640, A1. doi:10.1051/0004-6361/202038192
	
	\bibitem[Cantat-Gaudin \& Anders(2020)]{cant20a} Cantat-Gaudin, T. \& Anders, F.\ 2020, \aap, 633, A99. doi:10.1051/0004-6361/201936691
	
	\bibitem[Chen et al.(2004)]{chen04} Chen, W.~P., Chen, C.~W., \& Shu, C.~G.\ 2004, \aj, 128, 2306.
	doi:10.1086/424855
		
	\bibitem[Dib et al.(2018)]{dib18} Dib, S., Schmeja, S., \& Parker, R.~J.\ 2018, \mnras, 473, 849. doi:10.1093/mnras/stx2413	
		
	\bibitem[Hetem \& Gregorio-Hetem(2019)]{hete19} Hetem, A. \& Gregorio-Hetem, J.\ 2019, \mnras, 490, 2521. doi:10.1093/mnras/stz2698
	
	\bibitem[Hu et al.(2021a)]{hu21a} Hu, Q., Zhang, Y., Esamdin, A., et al.\ 2021, \apj, 912, 5. doi:10.3847/1538-4357/abec3e
	
	\bibitem[Hu et al.(2021b)]{hu21b} Hu, Q., Zhang, Y., \& Esamdin, A.\ 2021, \aap, 656, A49. doi:10.1051/0004-6361/202141460
		
	\bibitem[Kharchenko et al.(2009)]{khar09} Kharchenko, N.~V., Berczik, P., Petrov, M.~I., et al.\ 2009, \aap, 495, 807. doi:10.1051/0004-6361/200810407
	
	\bibitem[Kruijssen(2012)]{krui12} Kruijssen, J.~M.~D.\ 2012, \mnras, 426, 3008. doi:10.1111/j.1365-2966.2012.21923.x
	
	\bibitem[Lada et al.(1984)]{lada84} Lada, C.~J., Margulis, M., \& Dearborn, D.\ 1984, \apj, 285, 141. doi:10.1086/162485
	
	\bibitem[Lada \& Lada(2003)]{lada03} Lada, C.~J. \& Lada, E.~A.\ 2003, \araa, 41, 57. doi:10.1146/annurev.astro.41.011802.094844
	
	\bibitem[Lindegren et al.(2018)]{lind18} Lindegren, L., Hern{\'a}ndez, J., Bombrun, A., et al.\ 2018, \aap, 616, A2. doi:10.1051/0004-6361/201832727
	
	\bibitem[Luri et al.(2018)]{luri18} Luri, X., Brown, A.~G.~A., Sarro, L.~M., et al.\ 2018, \aap, 616, A9. doi:10.1051/0004-6361/201832964
		
	\bibitem[Millman \& Aivazis(2011)]{mill11} Millman, K.~J. \& Aivazis, M.\ 2011, Computing in Science and Engineering, 13, 9. doi:10.1109/MCSE.2011.36
	
	\bibitem[Miyamoto \& Nagai(1975)]{miya75} Miyamoto, M. \& Nagai, R.\ 1975, \pasj, 27, 533
	
	\bibitem[Meingast et al.(2021)]{mein21} Meingast, S., Alves, J., \& Rottensteiner, A.\ 2021, \aap, 645, A84. doi:10.1051/0004-6361/202038610
	
	\bibitem[McMillan et al.(2007)]{mcmi07} McMillan, S.~L.~W., Vesperini, E., \& Portegies Zwart, S.~F.\ 2007, \apjl, 655, L45. doi:10.1086/511763
	
	\bibitem[Moeckel \& Bonnell(2009)]{moec09} Moeckel, N. \& Bonnell, I.~A.\ 2009, \mnras, 400, 657. doi:10.1111/j.1365-2966.2009.15499.x
	
	\bibitem[Navarro et al.(1997)]{nava97} Navarro, J.~F., Frenk, C.~S., \& White, S.~D.~M.\ 1997, \apj, 490, 493. doi:10.1086/304888
	
	\bibitem[Netopil et al.(2022)]{neto22} Netopil, M., Oralhan, {\.I}. A., {\c{C}}akmak, H., et al.\ 2022, \mnras, 509, 421. doi:10.1093/mnras/stab2961
	
	\bibitem[Nilakshi et al.(2002)]{nila02} Nilakshi, Sagar, R., Pandey, A.~K., et al.\ 2002, \aap, 383, 153. doi:10.1051/0004-6361:20011719
	
	\bibitem[Oort(1979)]{oort79} Oort, J.~H.\ 1979, \aap, 78, 312
		
	\bibitem[Pang et al.(2021)]{pang21} Pang, X., Li, Y., Yu, Z., et al.\ 2021, \apj, 912, 162. doi:10.3847/1538-4357/abeaac
	
	\bibitem[Santos et al.(2005)]{sant05} Santos, J.~F.~C., Bonatto, C., \& Bica, E.\ 2005, \aap, 442, 201. doi:10.1051/0004-6361:20053378
		
	\bibitem[Smith \& Eichhorn(1996)]{smit96} Smith, H. \& Eichhorn, H.\ 1996, \mnras, 281, 211. doi:10.1093/mnras/281.1.211
	
	\bibitem[Tarricq et al.(2022)]{tarr21} Tarricq, Y., Soubiran, C., Casamiquela, L., et al.\ 2022, \aap, 659, A59. doi:10.1051/0004-6361/202142186
	
	\bibitem[Taylor(2005)]{tayl05} Taylor, M.~B.\ 2005, Astronomical Data Analysis Software and Systems XIV, 347, 29
	
	\bibitem[Zhai et al.(2017)]{zhai17} Zhai, M., Abt, H., Zhao, G., et al.\ 2017, \aj, 153, 57.
	doi:10.3847/1538-3881/153/2/57
	
	\bibitem[Zhang et al.(2020)]{zhan20} Zhang, Y., Tang, S.-Y., Chen, W.~P., et al.\ 2020, \apj, 889, 99. doi:10.3847/1538-4357/ab63d4 
	
	
	
	
	
	
\end{thebibliography}
\end{document}